\documentclass[pra,superscriptaddress,showpacs,twocolumn]{revtex4}
\usepackage{amsmath}
\usepackage{amstext}
\usepackage{latexsym}
\usepackage{graphicx}
\usepackage{amsfonts}

\begin{document}
\title{Test of nonlocality for a continuos-variable state based
on arbitrary number of measurement outcomes}

\author{W. Son}
\affiliation{School of Mathematics and Physics, The Queen's
University, Belfast BT7 1NN, United Kingdom}
\affiliation{The School
of Physics and Astronomy, University of Leeds, Leeds, LS2 9JT,
United Kingdom}
\author{{\v C}. Brukner}
\affiliation{Institut f\"ur Experimentalphysik, Universit\"at Wien, Boltzmanngasse 5,
A--1090 Wien, Austria}\affiliation{Institut f\"ur Quantenoptik und Quanteninformation,
\"Osterreichische Akademie der Wissenschaften, Boltzmanngasse 3, A--1090 Wien, Austria}
\author{M. S. Kim}
\affiliation{School of Mathematics and Physics, The Queen's
University, Belfast BT7 1NN, United Kingdom}

\date{\today}

\begin{abstract}

We propose a scheme to test Bell's inequalities for an arbitrary
number of measurement outcomes on entangled continuous variable (CV)
states. The Bell correlation functions are expressible in terms of
phase-space quasiprobability functions with complex ordering
parameters, which can experimentally be determined via local
CV-qubit interaction. We demonstrate that CV systems can give higher
violations of these Bell's inequalities than of the ones developed
for two-outcome observables.

\end{abstract}
\pacs{03.65.Ud, 03.67.-a, 42.50.-p, 42.50.Dv, 42.50.Pq, 85.25.Dq} \maketitle

Triggered by research in the foundations of quantum mechanics, in
more recent years, new ways of communication and computation were
discovered, which are based on quantum laws and can outperform their
classical counterparts. From a conceptual point of view one
distinguishes quantum information theory based on discrete and
continuous variable (CV) quantum systems.

On practical side the use of CVs has an advantage in efficient
implementation of the essential steps in quantum information
protocols, such as preparing, manipulating and measuring of
continuous quadrature amplitudes of the quantized electromagnetic
field. CV entanglement can be efficiently produced using squeezed
light and measured by homodyne detection~\cite{Braunstein05}. On the
conceptual side the advantage of CV systems over qubits originates
in {\it infinite-dimensionality} of its Hilbert space. For example,
a single CV system may be mapped on a discrete system of arbitrary
dimension and hence can be used as a universal resource in the
quantum protocols that are based on higher-dimensional systems
(qudits)~\cite{Brukner03}.

The notion of entanglement first appeared explicitly in the
literature in 1935, in a CV setting of Einstein, Podolski and Rosen
(EPR)~\cite{EPR35}. They considered an entangled state (the
EPR-state)
\begin{equation}
|\mbox{EPR}\rangle \propto \int dx \mbox{ } |x\rangle_1 \otimes
|x-Q\rangle_2, \label{epr}
\end{equation}
of two particles $1$ and $2$ that have perfectly correlated
positions ($x_1\!-\!x_2\!=\!Q$) and momenta $(p_1\!+\!p_2\!=\!0)$,
but is unnormalized. Here $|x\rangle_1 \otimes |x-Q\rangle_2$
denotes a product state of the two particles. The EPR state can be
thought of as the limiting case of properly normalized two-mode
squeezed vacuum (TMSV) state for infinitely large squeezing.
Experimentally, it can be produced by a nondegenerate optical
parametric amplifier.

Bell argued that the original EPR state allows a local realistic
description in terms of position and momentum because its Wigner
function is positive everywhere and hence can serve as a classical
probability distribution for local hidden variables~\cite{Bell87}.
Ironically, it was demonstrated that the Wigner function provides
a direct evidence of the nonlocal character of the states~\cite{Banaszek99},
though it does not lead to the maximal violation of Bell's inequality for the EPR state
(the Cirelson's bound of $2\sqrt 2$ can not be reached).  However, there is no fundamental contradiction between the two statements as in \cite{Banaszek99}, the form of the Wigner function merely coincides with the correlation function for parity measurements rather than it is used as the probability function to calculate the correlation. The approach has been generalized by introducing
pseudo-generators of SU(d) algebra for CV systems~\cite{Brukner03,Chen02}.
It established mathematical equivalence between discrete
$d$-dimensional and CV systems, which in principle allows for stronger violation of Bell's inequalities, but its experimental realization is
far beyond the reach of present technology. Thus, as for proposals
for revealing the nonlocality of entangled CV states, the current
state-of-art suggests that there is a {\it tradeoff} between the
efficiency and feasibility of the nonlocality tests.

Is it possible to find efficient tests of quantum non-locality for
CV systems without compromising too much on their feasibility? In
this paper we propose a feasible scheme that allows one to test an
arbitrary entangled state of CV systems as if it were a state of a
discrete system of arbitrary dimension. This
allows to test Bell's inequalities for arbitrarily high-dimensional
systems~\cite{Collins02}.  The measured correlations revealing the nonlocality
between higher-dimensional systems are shown to have a natural
description in terms of phase-space quasiprobability
functions with {\it complex} ordering parameters.  We demonstrate
that CV systems can give higher violation for Bell's
inequalities of larger dimension than for dichotomic observables~\cite{Banaszek99}. The recent
advances in the field of experimental quantum-state manipulation and measurements are shown to provide feasible experimental schemes for measuring quantum correlation functions~\cite{Lutterbach97,Bertet02}.

{\it Generic Bell function for higher dimensional systems-} In what
follows we will use a generic form of the Bell function for a
bipartite $d$ dimension as obtained in Ref.~\cite{Son05}:
\begin{eqnarray}
  \label{eq:vgbo}
  \hat{\cal B} = \sum_{n=1}^{d-1} G_n \hat{J}_A^n \otimes
  \hat{J}_B^n + h.c.
\end{eqnarray}
where the $n$-th order higher spin annihilation operators $\hat{J}_A^n$ and
$\hat{J}_B^n$ are given as
$\hat{J}_A^n=\frac{1}{2}(\hat{A}_1^n+\omega^{n/2} \hat{A}_2^n)$ and
$\hat{J}_B^n=\frac{1}{2}(\hat{B}_1^n+\omega^{n/2} \hat{B}_2^n)$. In
the construction of the annihilation operators, the measurement
operators $\hat{A}_i$ and $\hat{B}_i$ represent the measurement on
the parties $A$ and $B$ with the measurement settings of $i$.
Eigenvalues of the operator $\hat{A}_i$ and $\hat{B}_i$, which
correspond to the measurement outcomes, are assigned
by one of complex values $1,\omega, \omega^2, \cdot\cdot\cdot,
\omega^{(d-1)}$ where $\omega=e^{\frac{2\pi i}{d}}$. It is notable
that the generic Bell function gives a distinction between quantum
and classical statistics when a proper choice of the arbitrary
complex coefficient $G_n$ is made \cite{Son05}.

Due to the cyclic structure of the eigenvalues, the Hermitian
conjugate of the measurement operators are given as
$(\hat{A}^{\dagger})^{n}=\hat{A}^{(d-n)}$. Together with the
Hermitian conjugated part, the Bell operator can be expanded by the
measurement operators $\hat{A}$ and $\hat{B}$. Thus, the expectation
value of the Bell operator is given as
\begin{eqnarray}
\label{eq:expect}
\langle \hat{\cal B} \rangle
&=&\frac{1}{4}\sum_{n=1}^{d-1}\Big[ f_n^+ \left(
\langle\hat{A}_1^n\hat{B}_1^n\rangle+
\omega^{n}\langle\hat{A}_2^n\hat{B}_2^n\rangle\right)\\
&&~~~~~~~~+ f_n^-
\left(\omega^{n/2}
\langle\hat{A}_1^n\hat{B}_2^n\rangle+
\omega^{n/2}\langle\hat{A}_1^n\hat{B}_2^n\rangle
\right)\Big]\nonumber
\end{eqnarray}
where $f_n^{\pm}\equiv (G_n\pm G_{d-n}^*)$. Each correlation
function $\langle\hat{A}_i^n\hat{B}_j^n\rangle$ is the function of
probabilities for the measurement outcomes. With an antisymmetric
match of the outcomes, we have
\begin{eqnarray}
\label{eq:correlation1}
\langle\hat{A}_i^n\hat{B}_j^n\rangle&=&
\sum_{k,l=0}^{d-1}\omega^{n(k-l)}P(A_i=k, B_j=l)\nonumber\\
&=&\sum_{m=0}^{d-1}\omega^{n m}P(A_i-B_j\doteq m)
\end{eqnarray}
where $P(A_i=k, B_j=l)$ is the probability that each party obtains
the measured outcomes $k$ and $l$ from the measurement $(A_i, B_j)$.
$P(A_i-B_j\doteq m)$ is the probability that the difference of the
outcomes $(k,l)$ is equal to $m$ modulo $d$. The second equality is
followed because $\omega^{k}$ has the same value for same $k$ modulo
$d$.

With a proper choice of $f_n^{\pm}$, the classical expectation value
of the Bell function in Eq. (\ref{eq:expect}), under the local
realistic theory, is bounded by a certain value that is violated by
a quantum state. Interestingly, Collins-Gisin-Linden-Massar-Popescu
(CGLMP) inequality \cite{Collins02} which is known as tight Bell
inequality~\cite{Masanes02} for a bipartite $d$-dimensional system
can also be derived from it. As a non-trivial simplest case, the
CGLMP function for $d=3$ is
\begin{eqnarray}
\label{eq:CGLMP}
I_3&=& [P(A_1-B_1\doteq0)-P(A_1-B_1\doteq1)]\nonumber\\
&& -[P(A_2-B_2\doteq0)-P(A_2-B_2\doteq2)]\\
&&+[P(A_1-B_2\doteq0)-P(A_1-B_2\doteq2)]\nonumber\\
&&+[P(A_2-B_1\doteq0)-P(A_2-B_1\doteq2)]\nonumber
\end{eqnarray}
which is equivalent to the Bell function (\ref{eq:expect}) with
$f^{+}_n =\frac{4}{1-\omega^n}$ and $f^{-}_n
=-\frac{4\omega^{n/2}}{1-\omega^n}$. Through its generalization, one
can show that, by setting the coefficient function $f^{\pm}_n$ as
\begin{eqnarray}
\label{eq:coefficient} f^{\pm}_n&=& \pm
\frac{4}{d}\sum_{k=0}^{d-1}\left(\frac{2
k}{d-1}-1\right)\omega^{-\left(k+\frac{3\pm 1}{4}\right)n}
\end{eqnarray}
the Bell function in Eq. (\ref{eq:expect}) becomes CGLMP function
for arbitrary $d$ \cite{footnote}.  The function is proven to be
bounded by maximum value $2$ for any classical expectation value,
{\it i.e.} $\langle \hat{\cal B}\rangle\leq2$. In general, an
arbitrary correlation can be considered with a choice of
$f_n^{\pm}$. However, in the rest of our paper, we will only
consider the CGLMP inequality for our nonlocality test  since it is
tight for a bipartite $d$-dimensional case. For the nonlocality
test, we investigate a way to measure the correlation
(\ref{eq:correlation1}).


{\it Measurement of the complex observable for a CV state-}
One can ask how the complex observable is measured in a real
experimental setup. It is possible if one measures the $d$-level
system by discriminating all the $d$ outcomes and assign the
corresponding complex values for the measured outcomes.
A CV state is possible to be measured by a $d$ outcome measurement,
if there is a measurement that makes an arbitrary high dimensional
system projected onto a $d$-dimensional subspace \cite{Brukner03}.
For this purpose, we introduce a $d$-modulo observable
\begin{equation}
\label{eq:observable}
\hat{O}_{\alpha}=\hat{D}(\alpha)\sum_{n=0}^{\infty}e^{2i n
\pi/d}|n \rangle\langle n| \hat{D}^{\dagger}(\alpha),
\end{equation}
where the displacement operation
$\hat{D}(\alpha)=\exp(\alpha\hat{a}-\alpha^*\hat{a}^{\dagger})$
\cite{Cahill69}, with bosonic annihilation $\hat{a}$ and creation
$\hat{a}^\dag$ operators, is a feasible unitary operation for a CV
state \cite{Banaszek99}.  The observable (\ref{eq:observable}) maps
every possible number states $|n\rangle$ into $d$-modulo complex
numbers in a unit circle and its expectation value for a state of
density operator $\hat{\rho}$ is
\begin{equation}
\langle\hat{O}_{\alpha}\rangle =\sum_{n=0}^{\infty}
\left(\frac{z+1}{z-1}\right)^{n}\langle
n|\hat{D}^{\dagger}(\alpha)\hat{\rho}\hat{D}(\alpha)|n\rangle
\label{eq:10}
\end{equation}
where $z=-i\cot\pi/d$.  It is interesting to note that the
expectation value of the complex observable (\ref{eq:observable}) is
proportional to the quasiprobability function $W(\alpha,z)$ whose
ordering parameter can be shown to be
a complex number $z$ (see Ref.~\cite{Cahill69}):
\begin{equation}
\langle\hat{O}_{\alpha}\rangle
= \frac{\pi (1-z)}{2} ~W(\alpha,z).
\label{eq:11}
\end{equation}
One noticeable difference from a usual quasiprobability function
frequently considered in quantum optics is that `$z$' is complex
instead of real.  When $z=0$, the case reduces to the dichotomic
measurement~\cite{Banaszek99} and the expectation value for the
observable is proportional to the Wigner function~\cite{Cahill69}.
The expectation value (\ref{eq:10}) can be reconstructed by optical
tomography~\cite{Vogel89}. {\it Au contraire}, the observable
(\ref{eq:observable}) has also been measured
directly~\cite{Lutterbach97,Bertet02}  for $d=2$, using atomic
interferometry and dispersive atom-field interaction.  A two-level
atom (a qubit device) initially prepared in its ground state passes
through three interaction zones with the centre zone being the
interaction with the field whose even or odd parity is to be
measured.  In the first and last interaction zones, the atom
interacts with external fields for atomic state rotation,
$\hat{R}_i={1\over\sqrt{2}}\openone + {i\over
\sqrt{2}}(\sin\varphi_i\hat\sigma_x+\cos\varphi_i\hat\sigma_y)$.
Here, $\hat\sigma_{x,y,z}$ are Pauli spin operators and $\varphi_i$
($i=1,3$ represents the interaction zone) is determined by the phase
of the external field.  In the center zone, the evolution is
arranged for the atom-field evolution operator is described by
$\hat{U}=\exp[i{\phi\over
2}\openone+i(\hat{n}+{1\over2})\hat{\sigma}_z)]$, where
$\hat{n}=\hat{a}^\dag\hat{a}$ and $\phi$ is the interaction
parameter.  After passing through the interaction zones, the atom is
measured in its excited or ground state.  Considering only the
relevant terms, the field-atom state at the atomic measurement
is~\cite{Lutterbach97}
\begin{equation}
\label{extra} |g\rangle\langle g|\otimes \hat{M}_-\hat\rho\hat{M}_-
+  |e\rangle\langle e|\otimes \hat{M}_+\hat\rho\hat{M}_ +,
\end{equation}
where $\hat{M}_\pm=1\pm\mbox{e}^{i(2\hat{n}+1)\phi+i\eta}$,
$\eta=\varphi_2-\varphi_1$ and $|g, e\rangle$ represent the atomic
ground and excited states. For $\phi=-\eta=\pi/2 $, we find that the
field in an even (odd) photon-number state is signaled by the atom
measured in the excited (ground) state.  Recently, this has been
realized by an experiment \cite{Bertet02} to measure the parity for
a displaced Fock state prepared in a cavity.  The displacement
operation, $\hat{D}(\alpha)$, was performed by driving the cavity by
an external field.  The atom-field evolution, $\hat{U}$, was
implemented by atom-field interaction at non-zero detuning $\delta$
for the time  $t_{2}=\frac{\pi\delta}{4 \Omega^2}$ where $\Omega$ is
the Rabi frequency.   With use of the relation (\ref{eq:11}), the
Wigner function was measured for the Fock state with deep negativity
and detailed fringes. The detection efficiency to discriminate the
atomic state is highly efficient while there is a non-negligible
chance to miss atoms by a detector. However missing atoms is not a
problem in our non-demolition measurement setup and the measurement
device can be activated only at a right interaction time. An error
in the atomic interaction time is due to the atomic velocity
fluctuation around 3\%.

Even though the experimental model we have here is based on cavity
quantum electrodynamics, this can be done by any physical setup
which realizes a type of Ramsey interferometer with the dispersive
CV field-qubit interaction.  For example, direct measurements of the
quantum state have been proposed for the vibrational mode of a
trapped ion and an electromagnetic field in a superconducting
transmission line~\cite{Spiller05}.

Since the above scheme is a nondemolition measurement, it can be
repeated to perform further measurements.  By repeating the
procedure for a time $t_4={1\over2}t_2$, we can project the field
onto the subspaces spanned by $4n$ and $4n+2$, once the first atom
projected the field onto the even photon number state (by $4n+1$ and
$4n+3$ for the odd photon number state). The method can be applied
successively for the $d$ outcomes if $d$ is a power of two but,
unfortunately, it does not seem straightforward to generalize this
scheme to any measurement outcomes.

 If we are experimentally
conditioned to have only the expectation value of the observable
(instead of the individual measurement outcomes), (\ref{eq:10}), we
can use the scheme above for any number of measurement outcomes.
Setting the interaction time $t_d={2\over d} t_2$ and atoms to
interact with the field, then measure the difference
($\Delta_\eta=P_e-P_g$) in the probabilities of the atoms being in
the excited ($P_e$) and the ground ($P_g$) states.  Using
Eq.~(\ref{extra}), we can prove that the difference corresponds to
the expectation value:
\begin{equation}
\Delta_{\eta} = -\mbox{Re}\left[e^{i(\eta+\frac{\pi}{d})}
\langle\hat{O}(\alpha)\rangle\right]. \label{eq:measure}
\end{equation}
If we remind that the expectation value is complex-valued,
it is required to measure both its real and imaginary
parts, which is possible by changing $\eta$ dependent upon
the number of outcome $d$ as one can find in Eq.(\ref{eq:measure}).

{\it Measuring the correlation for Bell violation -} In order to
test the violation of the CGLMP inequality (\ref{eq:expect}), we
need to measure the correlation
$\langle\hat{A}_i^n\hat{B}_j^n\rangle$. As we restrict our system to
a CV state and the unitary operation to displacement operation
$\hat{D}(\alpha)$, the correlation function which has to be measured
is $\langle\hat{O}^n_{\alpha_i}\hat{O}^{n
\dagger}_{\beta_j}\rangle$. After an arrangement, one can find that
the correlation function corresponds to the two-mode
quasiprobability function
\begin{eqnarray}
\langle\hat{O}^n_{\alpha_i}\hat{O}^{n \dagger}_{\beta_j}\rangle&=&
\frac{\pi^2(1-z_n^2)}{4} W(\alpha_i,\beta_j,z_n)
\label{ext1}
\end{eqnarray}
where
\begin{eqnarray}
\label{eq:tmquasi}
 &&W(\alpha_i, \beta_j, z_n)=\sum_{m,l=0}^{\infty}
\left(\frac{z_n+1}{z_n-1}\right)^{m-l}\\
&&~~~~~~~~~~~~\times \langle m,
l|\hat{D}^{\dagger}(\alpha_i)\hat{D}^{\dagger}(\beta_j)
\hat{\rho}\hat{D}(\alpha_i)\hat{D}(\beta_j)|m,l\rangle\nonumber
\end{eqnarray}
and $z_n=\frac{\omega^n+1}{\omega^n-1}= -i\cot\frac{n\pi}{d}$.

To measure the correlation for the outcomes $d=2^\ell$, we let
$\ell$ number of atoms interact with each mode of the field at
appropriate interaction times then assign the complex values to each
measurement outcomes as in (\ref{eq:tmquasi}). On the other hand,
the expectation value (\ref{eq:tmquasi}) itself can also be measured
for any number of outcomes with $t_{d/n} = {2n\over d} t_2$. For the
case, the local atom-field interactions are adjusted with local
parameters $\eta_a$ and $\eta_b$, resulting in
\begin{eqnarray}
\Delta_{\eta_a,\eta_b}
=\frac{1}{2}\mbox{Re}\left[e^{i(\eta_a+\eta_b+\frac{2\pi n}{d})}
\langle\hat{O}^n_{\alpha_i}\hat{O}^n_{\beta_j}\rangle+e^{i(\eta_a-\eta_b)}
\langle\hat{O}^n_{\alpha_i}\hat{O}^{n
\dagger}_{\beta_j}\rangle\right] \nonumber
\end{eqnarray}
where $\Delta_{\eta_a,\eta_b}= P_{ee}+P_{gg}-P_{eg}-P_{ge}$ and
$P_{ee}$, for example, is the probability that the both atoms are in
the excited states. After fixing $\eta_b=-\pi n/d$, one measures
$\Delta_{\eta_a, -{\pi n\over d}}$.  Then readjusting $\eta_b=-(2
n+d)\pi/2d$, measures  $\Delta_{\eta_a, -{(2 n+d)\pi\over 2d}}$. The
addition of the two is related to the correlation function:
\begin{eqnarray}
\Delta_{\eta_a,-\frac{\pi n}{d}}+\Delta_{\eta_a, -\left(\frac{\pi
n}{d}+\frac{\pi}{2}\right)} =\mbox{Re}\left[e^{i(\eta_a+\frac{\pi
n}{d}+\frac{\pi}{2})} \langle\hat{O}^n_{\alpha_i}\hat{O}^{n
\dagger}_{\beta_j}\rangle\right].
\end{eqnarray}
We can immediately see that pre-setting $\eta_a=-(2n+d){\pi \over
2d}$ ($\eta_a=(n-d){\pi\over d}$), we obtain real (imaginary) part
of the correlation function, which is interesting to note.

\begin{figure}[ht]
{\bf (a)}\hskip4cm{\bf (b)}
\begin{center}
\centerline{\includegraphics[width=1.65in]{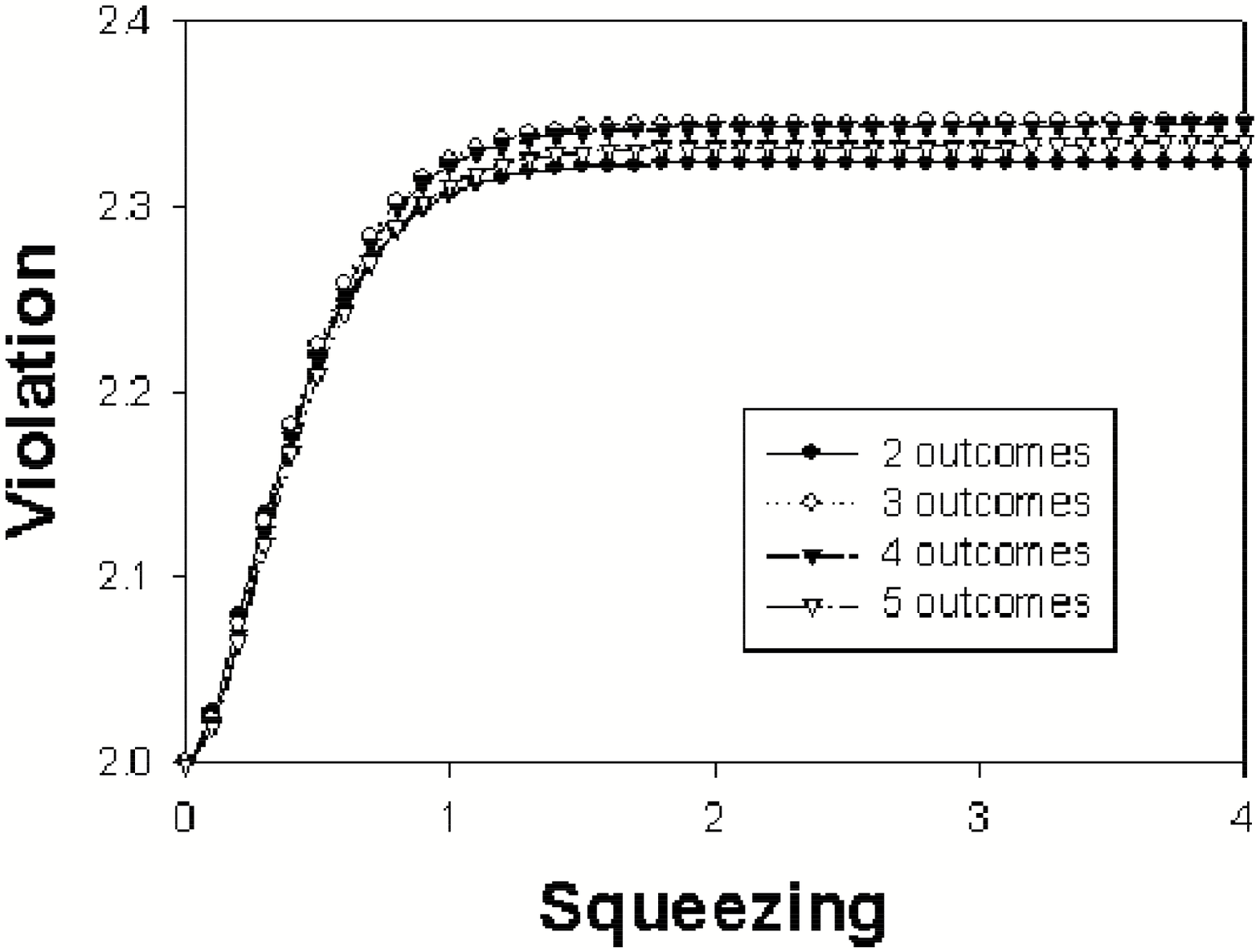}
\includegraphics[width=1.55in]{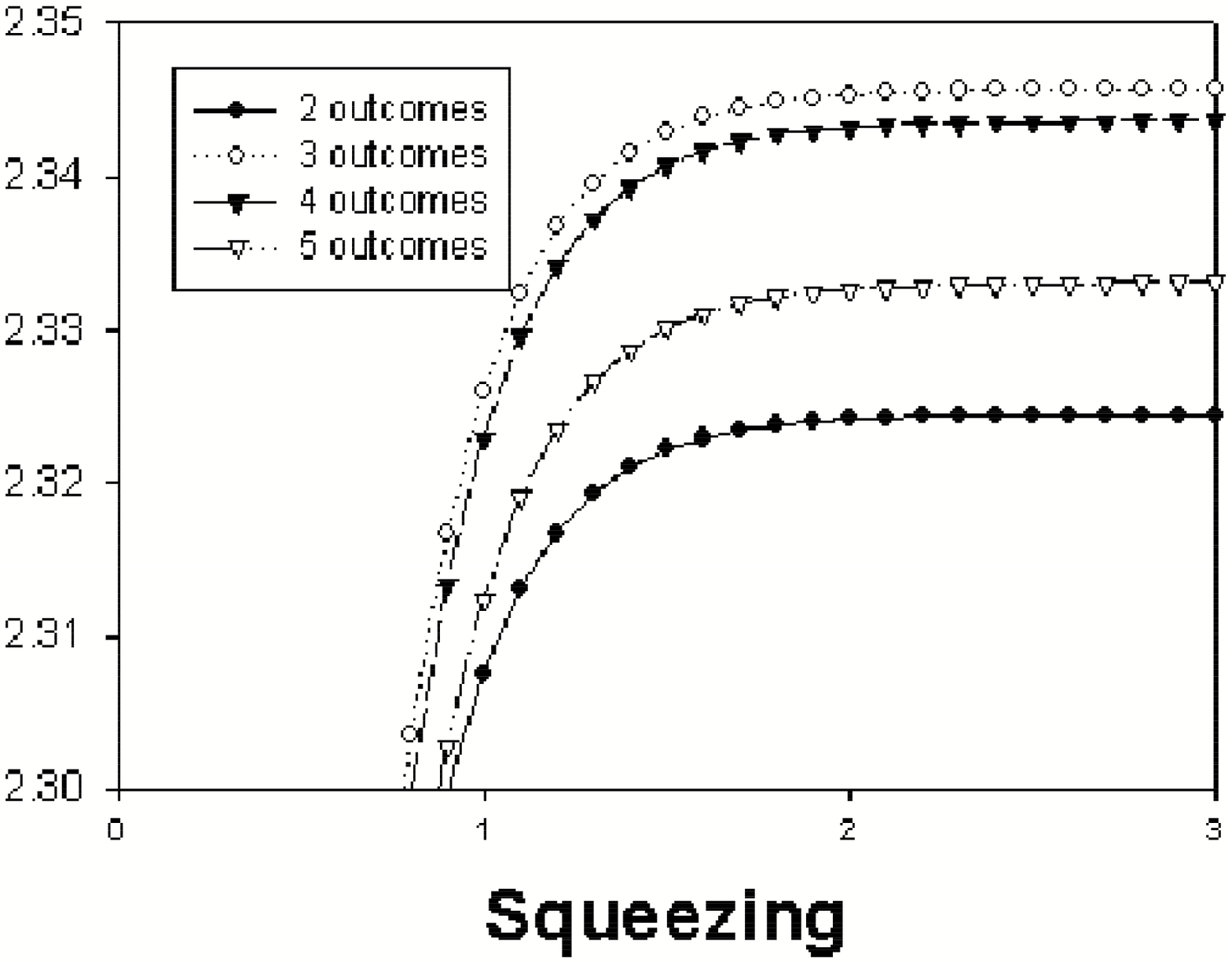}}
\end{center}
\caption{(a) Violation of local realistic bound 2 with complex
observable for a two-mode squeezed state. Re-scaled figure (b)
clearly shows the difference between the strength of violation for
the different numbers of measurement outcome.} \label{fig:violation}
\end{figure}

{\it Violation of CGLMP inequality for TMSV state- } A TMSV state
$|\psi\rangle$ is an entangled CV state: $|\psi\rangle =
\sum_{n=0}^{\infty} \frac{(\tanh r)^n}{\cosh r}|n,n\rangle_{A,B}$,
where $r$ is the squeezing parameter. It is well-known that when
squeezing parameter $r$ goes infinity, the TMSV state approaches to
the EPR state (\ref{epr}). The complex quasiprobability function for
the TMSV state is
\begin{eqnarray}
W_{\mbox{\tiny TMSV}} (\alpha_i,\beta_j,z_n;r) = \frac{4}{\pi^2
(1-z_n^2)}~~~~~~~~~~~~~~~~~~~~~~~~~~~~~~\\
\times\exp\left[-2\frac{A|\alpha_i|^2+A^*|\beta_j|^2+\sinh 2r
(\alpha_i\beta_j+\alpha_i^*\beta_j^*)}{(1-z_n^2)}\right]\nonumber
\end{eqnarray}
where $A\equiv\cosh2r +z_n$. With the complex quasiprobability
function, we test the violation of CGLMP inequality for the TMSV
state. The CGLMP is bounded by 2 for any local realistic model. The
violation of CGLMP inequality is plotted in Fig.~\ref{fig:violation}
as the function of squeezing parameter for different numbers of
measurement outcome, {\it i.e} different $d$. The function is
plotted after the linear optimization which maximizes the function
for the local parameters $\alpha_1$, $\alpha_2$, $\beta_1$ and
$\beta_2$. It shows that TMSV state always violates the constraint
given by local realistic model irrespective of the amount of
squeezing and the number of outcomes. The amount of violation is in
the order of 3 $>$ 4 $>$ 5 $>$ 2 outcomes in this case. The reason
that the violation does not increase monotonously by increasing the
number of measurement outcome is due to the restricted access to
possible measurement  (displacement operation instead of the full
SU($d$) unitary operation).

{\it Conclusions -} In many of the $d$-outcome nonlocality test, the
observable can be non-Hermitian giving a complex number as a
measurement outcome. It is not obvious how to construct an
experiment for such a measurement.  We have proposed, for the first
time, an efficient and feasible scheme to test Bell's inequality for
an arbitrary number of measurement outcomes on CV systems. The
nonlocal correlation functions directly corresponds to
quasiprobability functions with complex ordering parameter in phase
space. This relation makes the experimental realization for the
$d$-outcome nonlocality test feasible. We have shown that entangled
CV systems can violate this Bell inequality beyond the limits
obtained in the tests of standard Bell's inequalities for dichotomic
observables.


We acknowledge support of the UK EPSRC, Austrian Science Foundation
(FWF) Project SFB 1506, European Commission (RAMBOQ), KRF
(2003-070-C00024) and the British Council in Austria.



\begin{thebibliography}{99}
\bibitem{footnote} Each correlation term in Eq. (\ref{eq:correlation1})
can be written as $\frac{1}{4}\sum_{n=1}^{d-1} f_n^{\pm}\omega^{n
\gamma} \langle\hat{A}^n\hat{B}^n\rangle=\sum_{m=0}^{d-1}
g^{\pm}_{(m+\gamma)} P(A-B\doteq m)$ with
$g^{\pm}_{(m+\gamma)}=\pm\frac{1}{d}\sum_{k,n=0}^{d-1}\Big(\frac{2k}{d-1}-1\Big)
\omega^{(m+\gamma-k-\frac{3\pm 1}{4})n}$. Evaluating one of $g$
functions, for example $g^+_{(m+1)}$, we have $g^+_{(m+1)}
=\frac{2m}{d-1}-1$. Similarly, one can find $g^{+}_{m}$ and
$g^{-}_{m+1/2}$ and with the given functions of $f^{\pm}_n$, it is
possible to check the equivalence of our Bell function and the CGLMP
one.
\end{thebibliography}
\end{document}